\documentclass[aps,prb,preprint,superscriptaddress]{revtex4-1}
\usepackage{graphicx}
\usepackage{amsmath}
\usepackage{amssymb}
\usepackage{mathtools}
\usepackage{xcolor}
\usepackage{braket}
\usepackage[colorlinks=true, citecolor=blue, urlcolor=blue, linkcolor=blue,bookmarks=false,hypertexnames=true]{hyperref} 
\usepackage{appendix}
\usepackage{array}
\usepackage[column=O]{cellspace}
\newcolumntype{P}[1]{>{\centering\arraybackslash}p{#1}}

\begin{document}
\graphicspath{}
\preprint{APS/123-QED}

\title{Superexchange Mechanism in Coupled Triangulenes Forming Spin-1 Chains}
\date{12 March, 2024}

\author{Yasser Saleem}
\affiliation{Institut f\"{u}r Physikalische Chemie, Universit\"{a}t Hamburg, Grindelallee 117, D-20146 Hamburg, Germany}

\author{Torben Steenbock}
\affiliation{Institut f\"{u}r Physikalische Chemie, Universit\"{a}t Hamburg, Grindelallee 117, D-20146 Hamburg, Germany}

\author{Emha Riyadhul Jinan Alhadi}
\affiliation{Institute of Physics, Faculty of Physics, Astronomy and Informatics, Nicolaus Copernicus University, Grudziadzka 5, 87-100 Toru\'n, Poland}

\author{Weronika Pasek}
\affiliation{Institute of Physics, Faculty of Physics, Astronomy and Informatics, Nicolaus Copernicus University, Grudziadzka 5, 87-100 Toru\'n, Poland}

\author{Gabriel Bester}
\affiliation{Institut f\"{u}r Physikalische Chemie, Universit\"{a}t Hamburg, Grindelallee 117, D-20146 Hamburg, Germany}

\author{Pawel Potasz}
\affiliation{Institute of Physics, Faculty of Physics, Astronomy and Informatics, Nicolaus Copernicus University, Grudziadzka 5, 87-100 Toru\'n, Poland}
\begin{abstract}

We show that the origin of the antiferromagnetic coupling in spin-1 triangulene chains, which were recently synthesized and measured by Mishra et al. Nature {\bf 598}, 287-292 (2021) originates from a superexchange mechanism. This process, mediated by inter-triangulene states, opens the possibility to control parameters in the effective bilinear-biquadratic spin model. We start from the derivation of an effective tight-binding model for triangulene chains using a combination of tight-binding and Hartree-Fock methods fitted to hybrid density functional theory results. Next, correlation effects are investigated within the configuration interaction method. 
Our low-energy many-body spectrum for $N_{\rm Tr}=2$ and $N_{\rm Tr}=4$ triangulene chains agree well with the bilinear-biquadratic spin-1 chain antiferromagnetic model when indirect coupling processes, and superexchange coupling between triangulene spins are taken into account. 


\vspace{6mm}

Corresponding author email: yassersaleem461@gmail.com
\end{abstract}
\maketitle



The antiferromagnetic (AFM) Heisenberg spin-1 chain exhibits intriguing properties, including symmetry-protected topological order and a gapped excitation spectrum known as the Haldane gap. \cite{HaldaneNonlinear, HALDANE1983464,affleck2004rigorous,Affleck_1989} Furthermore, in finite systems, characteristic spin-$S=\frac{1}{2}$ edge excitations emerge \cite{Kennedy1990ExactDO}. Affleck, Kennedy, Lieb, and Tasaki (1987) proposed the AKLT model to elucidate the origin of the Haldane gap, introducing an isotropic one-dimensional spin-1 model with a unique ground state that also confirms the presence of an energy gap to excited states \cite{affleck2004rigorous}. This model, characterized by an additional biquadratic exchange term, allows for a representation of the ground state in a valence-bond basis. The AFM Heisenberg spin-1 chain, along with models like AKLT, belongs to a broader class of bilinear-biquadratic (BLBQ) spin-1 chain models, which display diverse magnetic orderings dependent on the energetic ratio between linear and biquadratic exchange terms (see Refs. \onlinecite{Fath1995BLBQPhases,BLBQ_Cirac,BLBQphases_Lauchli} and references therein).


Spin models are effective models of the low-energy subspaces of a Fermionic system\cite{Rossier2022HubbardForSpin1,Henriques2023t3paper,Manalo2024}. They can be derived from other simplified models such as the Hubbard model, which is an approximation to the full interacting Hamiltonian containing all two-body Coulomb interactions. AFM coupling in spin models can be explained by the kinetic exchange mechanisms, a coupling between singly and doubly occupied sites by effective hopping. This process favours AFM ordering of electron spins, since electrons with opposite spins are able to freely hop from one site to the next without being repelled due to Pauli exclusion \cite{koch2012exchange}. Another exchange mechanism, known as superexchange has been  proposed to explain AFM order in transition metal oxides,  \cite{ANDERSON196399,Anderson1959SuperExchange, Moriya1960SuperExchange,koch2012exchange}. In that case, two cation orbitals are separated by an anion, as such, direct hopping between cations is quenched, while hopping between localized d-orbitals (located on the cation) occurs through an intermediate p-state (located on the anion)\cite{Anderson1959SuperExchange,rao1989transition,koch2012exchange,Tokura2000TMO,cox2010transition,Bhattacharya2014MagneticOxides}.

Recently, in 2021 Mishra et al. \cite{FaselNaturespinchain} observed signatures of fractional edge states, and gapped spin excitations in spin-1 triangulene chains. An AFM coupling between neighboring spins was described as a direct process in an effective Hubbard model with an enhanced next-next-nearest-neighbor (NNNN) hopping commonly labelled $t_3$ in literature\cite{Henriques2023t3paper,FaselNaturespinchain,Mart2023Theorypaper}. While that model captures major physical properties of the system, we show that the microscopic mechanism of the AFM coupling is an indirect superexchange process that includes states localized on the connection of triangulene molecules. This opens up  opportunities to control paramaters of the spin model in a variety of ways.

In our work, we start from a combination of tight-binding (TB) and  Hartree Fock (HF) methods and fit the results to density functional theory (DFT), in order to derive an effective single-particle model of triangulene chains that includes the presence of valence band electrons. The obtained spatial variation of TB hopping integrals takes into account geometric effects. We include correlations within the complete active space (CAS) using the configuration interaction (CI) method. We analyze the low-energy many-body spectrum as a function of CAS size and compare it with experimental results from Ref. \onlinecite{FaselNaturespinchain}. We
find that the direct kinetic exchange contribution is too small to explain the spin-1 AFM model spectrum and the exchange mechanism that dominates the value of the coupling constant $J$ is due to superexchange. Finally, we fit the many-body spectrum to the BLBQ model for $N_{\rm Tr}=2$ and $N_{\rm Tr} =  4$ triangulene chains, confirming that at low-energies, the system behaves as a spin-1 chain, and allows us to extract effective spin parameters.

\begin{figure}[t]
    \centering
    \includegraphics[width=0.5\linewidth]{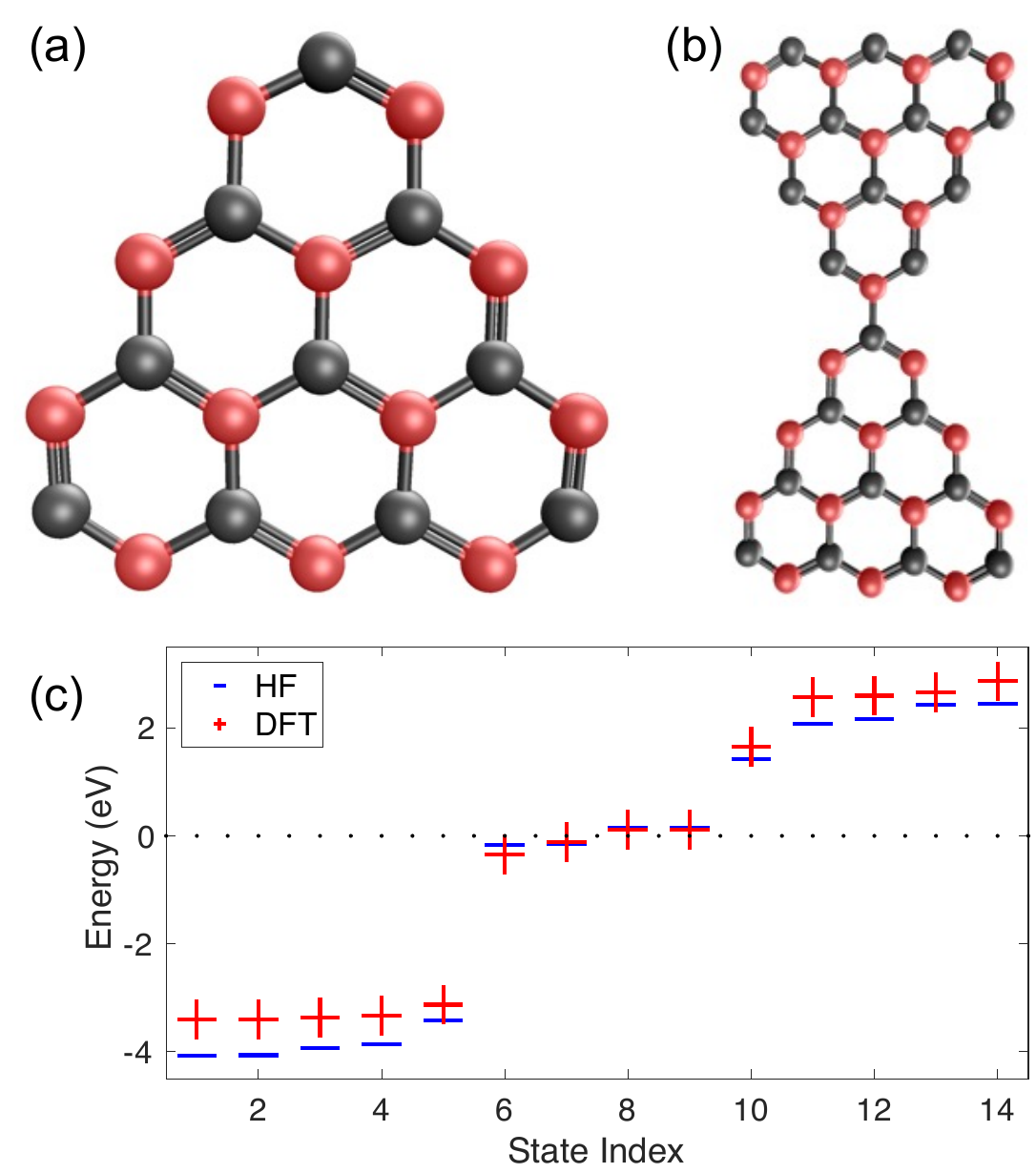}
    \caption{{} Geometrical representation of  (a) triangulene and (b) the triangulene dimer. The balls denote location of carbon atoms with the red balls corresponding to sublattice A and the gray balls corresponding to sublattice B. The sticks connecting carbon atoms refer to the alternating single and double bonds. In (c) we show the single-particle spectrum of the triangulene dimer for states near the Fermi level. The blue horizontal lines are HF results, while the red crosses correspond to DFT results obtained with the B3-LYP exchange--correlation functional. The dotted line corresponds to the Fermi level at half-filling}
   \label{fig:system}
\end{figure}

Triangular graphene quantum dots (TGQDs) with zigzag edges exhibit promising potential as building blocks for spin chains. This is because the single-electron spectrum for these quantum dots collapses to a degenerate shell at the Fermi level, a phenomenon resulting from the sublattice imbalance. Consequently, it is predicted that these quantum dots possess a magnetic spin-polarized ground state (GS) at half-filling\cite{Rossier2007,Ezawa2007,Ezawa2008,wang2008graphene,Wang2008NL,GuccluPRL2009,Potasz2010ZeroEnergy,potasz2012electronic,saleemquantumsimulator,Guclu2023AG}, a property consistent with Lieb's theorem\cite{LiebTheorems}. The degeneracy of this shell is proportional to the length of a single edge of the triangular structure \cite{Potasz2010ZeroEnergy}. Triangulene, which is the second smallest TGQD with zigzag edges, containing 22 carbon atoms (also referred to as Clar's hydrocarbon), is of particular interest in this work and is shown in Fig.~\ref{fig:system}(a). Triangulene has two degenerate states at the Fermi level and consequently possesses a spin-1 ground state. It was predicted to exist and attempted to be synthesized in 1953 by Clar and Stewart\cite{Clar1953}. Triangulene faced synthesis challenges due to the molecule's high reactivity arising from the two unpaired electrons. Overcoming this challenge more than 60 years later, Pavli{\v c}ek et al.\cite{pavlivcek2017synthesis} successfully synthesized triangulene in 2017. Following this milestone, noteworthy experimental work focused on the synthesis of triangulene as well as larger-sized TGQDs emerged \cite{mishra2021synthesis,Su2019Synthesis,mishra2019synthesis}.

The first step in constructing a synthetic spin-1 chain using triangulene involved combining two triangulene molecules to form a triangulene dimer (TD) as illustrated in Fig.~\ref{fig:system}(b). This was accomplished experimentally \cite{mishra2020collective,FaselNaturespinchain} and investigated theoretically \cite{Hongede2023NL,Henriques2023t3paper,Rossier2022HubbardForSpin1,ortiz2022theory} in recent years. To understand the single-particle properties of the TD, we employ a $p_z$ orbital TB model, akin to previous studies \cite{Henriques2023t3paper,Saleem_2019,graphenebook,Zarenia2011QDs}. One assumption in these TB models, is that that the hopping parameters between two sites depends solely on the distance between them. However, when graphene is cut into a finite flake, electron hopping can be dramatically modified especially along the edge. We correct the hoppings by solving the HF equation given by \cite{graphenebook,potasz2012electronic,saleemquantumsimulator}



\begin{equation}
\begin{split}
    H_{MF}&= \sum\limits_{i,l,\sigma}t_{il\sigma} c_{i\sigma}^\dag c_{l\sigma} \\
   &+\sum\limits_{i,l,\sigma}\sum\limits_{j,k,\sigma'}\left(\braket{ij|V|kl}-\braket{ij|V|lk}\delta_{\sigma\sigma'}\right)\\
   &\times\left(\rho_{jk\sigma'}-\rho^0_{jk\sigma'}\right)c^\dag_{i\sigma}c_{l\sigma} \\
   &=\sum\limits_{i,l,\sigma}\Tilde{t}_{il\sigma} c_{i\sigma}^\dag c_{l\sigma},
    \end{split}
     \label{eq:HFHam} 
    \end{equation}
where  $c^\dagger_{i,\sigma}/c_{i,\sigma}$ creates/annihilates a $p_z$ electron on site $i$ with spin $\sigma$.  {$t_{il\sigma}$} are the two-dimensional (2D) infinite system hopping elements. We take the following standard accepted values for {$t_{il\sigma}$}: $t=-2.8$ eV as the nearest-neighbor (NN) hopping, $t'=-0.1$ eV as the next-nearest-neighbor (NNN) hopping and $t_3=-0.07$ eV is the NNNN hopping \cite{Castroelectronicproperties}. $\rho^0_{jk\sigma'}$ are the density matrix elements for a 2D infinite graphene sheet obtained previously\cite{potasz2012electronic,graphenebook}.   $\rho_{jk\sigma'}$ are the density matrix elements for the finite system computed with respect to the HF ground state. The Coulomb matrix elements $\braket{ij|V|kl}$ are screened by a dielectric constant $\kappa = 3$ which gives good agreement with DFT calculations, as seen in Fig. \ref{fig:system}, see details of calculations in Supplemental Materials (SM) \cite{SM}.

\begin{figure*}[t]
    \centering
    \includegraphics[width=\linewidth]{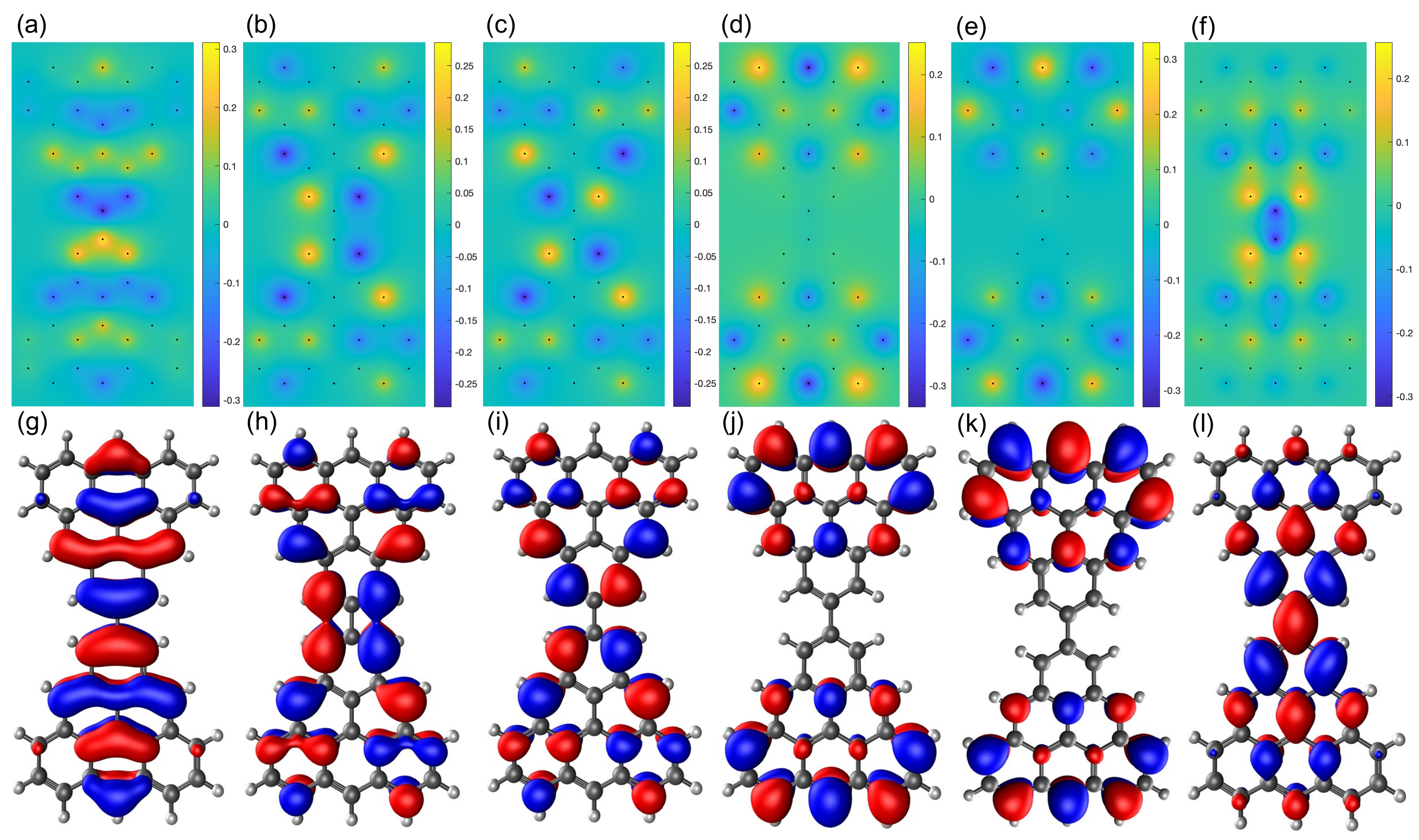}
    \caption{{} Hartree-Fock (a-f) and DFT (g-l)  wavefunctions for the six states located closest the the Fermi level at half-filling. (a-f) and (g-l) corresponds to the states 5-10 shown in Fig.~\ref{fig:system}(c).}
   \label{fig:2trianguleneWFcomparison}
\end{figure*}

For the TD shown in Fig.~\ref{fig:system}(b), there are $N_{\rm s}=44$ atomic sites, and in the single-orbital model here, this corresponds to $N_{\rm e}=44$ electrons for the charge neutral system. It is convenient for the CI calculations that will follow, to remove four electrons from the TD, similar to what was done in previous work \cite{potasz2012electronic,GuccluPRL2009,saleemquantumsimulator} and add them back when performing CI calculations. 
Therefore, we self-consistently solve the HF equation, as represented in Eq.~\ref{eq:HFHam}, for $N_\uparrow=20$ spin-up and $N_\downarrow=20$ spin-down $p_z$ electrons, which is justified by noticing the presence of the energy gap between the lowest $N_{\rm st}=20$ energy states and four degenerate shell states, seen in the tight-binding spectrum \cite{Henriques2023t3paper}. We obtain HF quasiparticle orbitals and compare them with the results from DFT calculations for a +4 cation TD. Fig.~\ref{fig:system}(c)  show agreement between DFT and HF energy levels, and Fig.~\ref{fig:2trianguleneWFcomparison} shows agreement between their corresponding wavefunctions. We use one fitting parameter between the two calculations, which is a static screening constant $\kappa=3$, which reduces the strength of Coulomb interaction by $1/\kappa$. Examination of Fig.~\ref{fig:system}(c) reveals the presence of four distinct states in close proximity to the Fermi level, and a substantial energy gap separates these states from others. The wavefunctions of states No. 5 (the first state below the degenerate shell, Fig.~\ref{fig:2trianguleneWFcomparison}(b) and (g)) and No. 10 (the first state above the degenerate shell, Fig.~\ref{fig:2trianguleneWFcomparison}(f) and (l)), are localized at the connection between the two triangulenes.
The wavefunctions of four degenerate shell states significantly differs from the wavefunctions of two degenerate states from an isolated triangulene as illustrated in Fig.~\ref{fig:2trianguleneWFcomparison}(b-e) for the HF calculations and (h-k) for DFT calculations (compared with the Fig. 1 inset in SM \cite{SM}). In particular, after HF self-consistent calculations, the degenerate shell states No. 6 (Fig.~\ref{fig:2trianguleneWFcomparison}(b) and (h)) and No. 7 (Fig.~\ref{fig:2trianguleneWFcomparison}(c) and (i)) strongly hybridize with the inter-triangulene states.  This implies that in order to analyze coupling between the spins of isolated trianglulenes, we must take into account at least the six states shown in Fig.~\ref{fig:2trianguleneWFcomparison}. 

\begin{figure}[t]
    \centering
    \includegraphics[width=\linewidth]{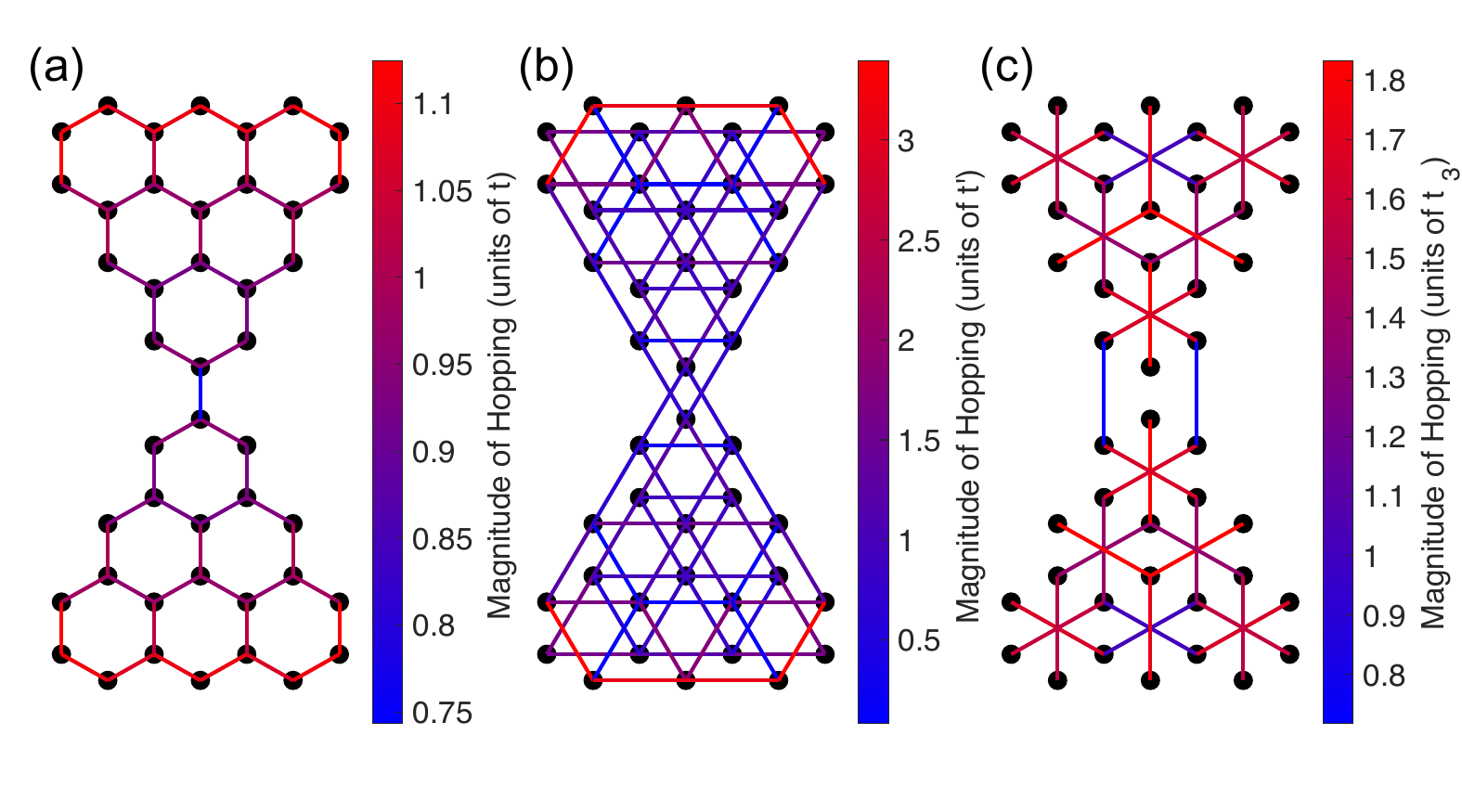}
    \caption{{}Charge neutral effective HF hopping elements for (a) nearest neighbor $t$ (b) next nearest neighbor $t'$ (c) next-next nearest neighbor $t_{\rm 3}$. The black dots correspond to the location of carbon atoms, and the line connecting them corresponds to the effective hopping between carbon atoms. The units for each color plot are measured with respect to their 2D infinite system hopping values. }
   \label{fig:HFHopping}
\end{figure}

As seen, in the last line of Eq.~\ref{eq:HFHam}, the HF equation reduces to a single effective hopping parameter $\Tilde{t}_{il\sigma}$ which accounts for finite-sized modifications of the TB hoppings. It is given explicitly as
 \begin{equation}
    \Tilde{t}_{il\sigma} = t_{il\sigma} +\sum\limits_{j,k,\sigma'}\left(\braket{ij|V|kl}-\braket{ij|V|lk}\delta_{\sigma\sigma'}\right)\left(\rho_{jk\sigma'}-\rho^0_{jk\sigma'}\right).
     \label{eq:HFHopping}
\end{equation}
It is apparent in Eq.~\ref{eq:HFHopping} that when $\rho_{jk\sigma'}=\rho^0_{jk\sigma'}$, the effective hopping parameter reduces to that of the extended system hopping parameters $t_{il\sigma}$. Furthermore, deviations of the hopping elements will largely depend on the density matrix elements, as such, to obtain effective hopping paramaters that more closely resemble those that should be used in a TB model, we recompute the density matrix $\rho_{jk\sigma'}$ at charge neutrality. Thus, all states below the dotted line in Fig.~\ref{fig:system}(c) are occupied. Fig.~\ref{fig:HFHopping} shows how the extended system hopping parameters are modified in HF for the TD. The line connecting two atoms represents the hopping between the two sites while the color represents the magnitude of the hopping. We see in Fig.~\ref{fig:HFHopping}(a) that in the center of the structure the effective NN hopping is almost equal to the 2D infinite system value $t$. Away from the center is where the largest modifications occur. There is an enhancement of the NN hopping element of about 10\% near the bottoms of each triangulene molecule, and a reduction of about 25\% on the connection between the two triangulenes. Similarly, NNN $t'$ hopping element in the center of each individual triangulene resembles that of the 2D infinite system, but a 300\% enhancement of the NNN hopping element along the bottoms of each triangulene molecule is observed, as seen in Fig.~\ref{fig:HFHopping}(b). Finally in Fig.~\ref{fig:HFHopping}(c), even though, we observe an overall enhancement of the NNNN $t_{\rm 3}$ hopping, we observe a reduction of it on the connection between the two triangulene molecules. 
This contradicts a strong  enhancement of $t_{\rm 3}$ in the effective Hubbard model considered in previous works \cite{FaselNaturespinchain,Henriques2023t3paper,Tran2017Thirdneighbor}. We notice that the renormalization of the hopping elements is dominantly attributed to lack of the charge uniformity brought upon by the edge effects of the TD.

\begin{figure*}[t]
    \centering
    \includegraphics[width=\linewidth]{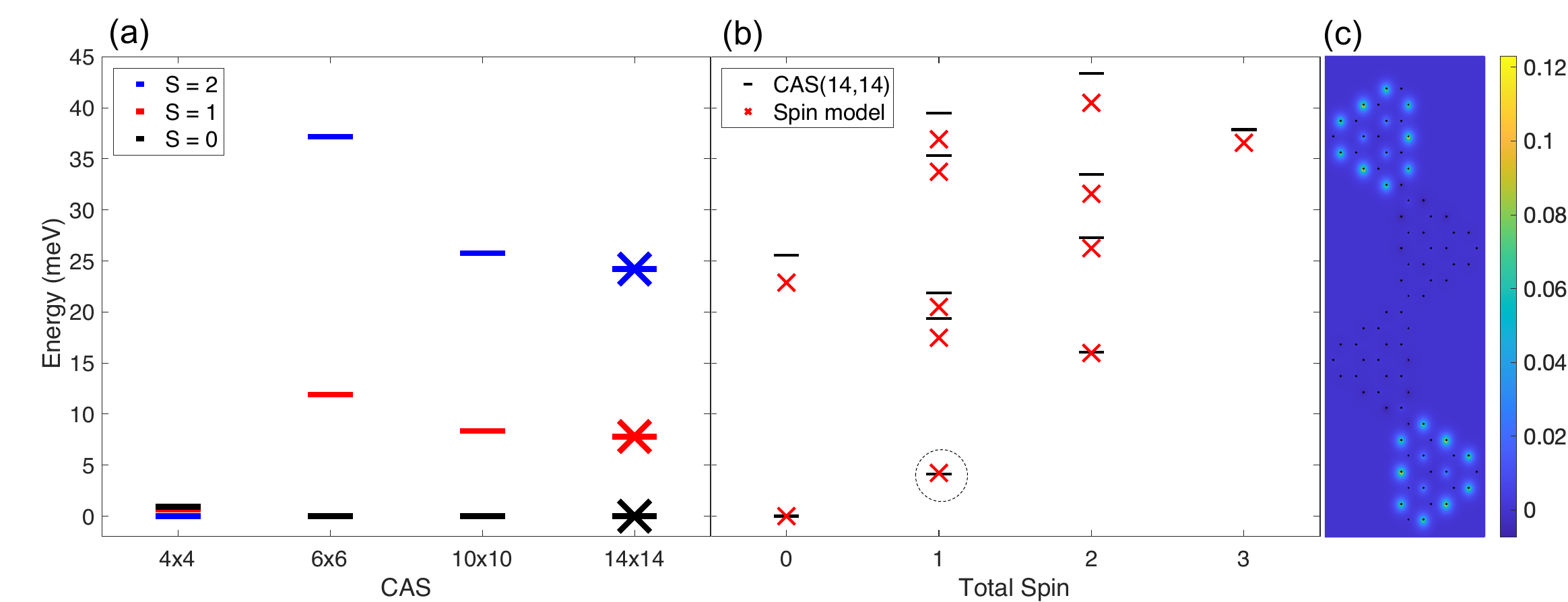}
    \caption{(a) Many-body spectrum for $N_{\rm Tr}=2$ triangulenes as a function CAS size. The x-axis denotes the number of states taken relative to the dotted line shown in Fig.~\ref{fig:system} (c), i.e. six states means taking 3 states above and 3 states below the dotted line.  The "x" data points overlaying the data for 14 states corresponds to the spin model results for $J=8.21$ meV and $\beta=0.016$. b) Comparison of the many-body spectrum for $N_{\rm Tr}=4$ triangulenes with corresponding BLBQ spin model results. In (c) we show the many-body spin density for the lowest energy triplet state for $N_{\rm Tr}=4$ indicated by a circle in (b).} 
   \label{fig:MBspectrumVsNumStates}
\end{figure*}

 In order to include the effects of correlations and obtain excited states of the TD we combine CI with the HF method. By rotating the many-body Hamiltonian to the basis of HF states one obtains \cite{saleemquantumsimulator, PfafflescreenedCI}
\begin{equation}
\begin{split}
 H&= \sum\limits_{p,\sigma}\epsilon^{HF}_{p\sigma}b_{p\sigma}^\dag b_{p\sigma}-\sum\limits_{p,q,\sigma} \tau_{pq\sigma}b_{p\sigma}^\dag b_{q\sigma}\\
 &+ \frac{1}{2}\sum\limits_{p,q,r,s,\sigma,\sigma'}\braket{pq|V|rs}b_{p\sigma}^\dag b_{q\sigma'}^\dag b_{r\sigma'} b_{s\sigma},
    \end{split}
\label{eq:ManyBodyHInBasisofHF}
\end{equation}
where $\epsilon^{HF}_{p\sigma}$ are the energies of the HF quasiparticle orbitals obtained by solving Eq.~\ref{eq:HFHam}.  $b^\dagger_{p,\sigma}/b_{p,\sigma}$ creates/annihilates an electron on the HF orbital $p$ with spin $\sigma$. $\tau_{pq\sigma}$ are the double counting corrections which arises to avoid double counting of the interactions that were already contained at the mean-field level \cite{saleemquantumsimulator, PfafflescreenedCI}. This effectively lowers the contribution of the last term in Eq.~\ref{eq:ManyBodyHInBasisofHF}. 

To solve the problem exactly for $N_{\rm Tr}=2$, one would need to include $N_{\rm e}=44$ electrons on $N_{\rm st}=44$ states, which is currently impossible to solve, due to an unmanageably large many-body Hilbert space. Thus, we solve the CI Hamiltonian given by Eq.~\ref{eq:ManyBodyHInBasisofHF} in the basis of configurations by considering CAS(4,4), CAS(6,6), CAS(10,10), and CAS(14,14) states, as shown in Fig.~\ref{fig:MBspectrumVsNumStates}(a), where we use the notation CAS($N_{\rm st}$,$N_{\rm e}$) with $N_{\rm st}/2$ taken above and $N_{\rm st}/2$ taken below the dotted line shown in Fig.~\ref{fig:system}(c). 

In the CAS(4,4) results, the ground state is ferromagnetic (FM) with $S=2$, and corresponds to a Hund's rule like filling of electrons on the degenerate shell consisting of four states.  This occurs when the Coulomb exchange amongst electrons on the shell of four states dominates compared to the effective AFM exchange coupling $J$ between the spin-1 quasiparticles.  When we include, into our active Hilbert space, the inter-triangulene HF states, $N_{\rm st}=6$, shown in Fig. \ref{fig:2trianguleneWFcomparison}(a) and \ref{fig:2trianguleneWFcomparison}(f), we observe the ground state transitions from FM to AFM. This AFM ground state, agrees with recent experiments \cite{FaselNaturespinchain}.  Electrons from the degenerate shell states can now effectively hop from one triangulene molecule to the neighboring one via the intermediate state. Thus AFM superexchange is needed in order to explain the spin-1 chain character recently observed experimentally\cite{FaselNaturespinchain}. Further increase of the CAS Hilbert space size only quantitatively corrects the values of the energy gaps between the singlet $S=0$, triplet $S=1$ and quintuplet $S=2$. The spin excitation gap measured experimentally \cite{FaselNaturespinchain} is comparable to  the results for CAS(14,14) ($14$ meV and $8$ meV, respectively) and could be improved by using a more advanced model for the dielectric screening.  

We fit our CAS(14,14) spectrum to the BLBQ model given by
\begin{equation}
 H_{BLBQ}= J\sum\limits_{i}\left[\Vec{S}_i\cdot\Vec{S}_{i+1}+\beta\left(\Vec{S}_i\cdot\Vec{S}_{i+1}\right)^2\right],
\label{eq:BLBQ}
\end{equation}
where $J$ is the bilinear exchange coupling constant between neighboring spin sites, and $\beta$ is it's biquadratic partner. By fitting to the BLBQ model, we extract the effective exchange terms $J=8.21$ meV and $\beta=0.016$. Finally, in Fig.~\ref{fig:MBspectrumVsNumStates}(b) we compare many-body spectra for $N_{\rm Tr}=4$ (we take all eight degenerate shell states and six inter-triangulene states in the calculations) and the corresponding energy spectrum of the spin Hamiltonian given by Eq. \ref{eq:BLBQ}. We find good agreement between energy spectra in both models for each total spin eigenstate proving the validity of spin-1 model description (with almost the same parameters as for TD case, the effective exchange terms are $J=8.29$ meV and $\beta=0.016$). Small differences in the excited state energies can be attributed to a finite Hilbert space. We note that calculations that include more HF states is beyond our numerical capabilities. Fig.~\ref{fig:MBspectrumVsNumStates}(c) shows the many-body spin density for the lowest energy triplet state, which resemble characteristic edge states in the Haldane phase. It is worth mentioning, that the excess of this spin density is similar to the one from the isolated triangulene (see SM\cite{SM}).

The AFM superexchange mechanism is usually derived using higher order perturbation theory. We notice that after self-consistent HF calculations, we already are in a basis of strongly hybridized states of the degenerate shell and the inter-triangulene states. This hybridization comes from interactions between electrons populating all filled states and virtual electrons on a degenerate shell, and thus can not be uncoupled by any rotation within the effective low-dimensional Hilbert space; one needs to include all the single-particle energy states, not only the four degenerate shell states (for $N_{\rm Tr}=2$) and inter-triangulene valence and conduction band states. This makes it difficult to explicitly identify the superexchange processes responsible for the antiferromagnetic coupling between two $S=1$ spins within perturbation theory. The degenerate states from isolated triangulene molecules are significantly different than the four HF degenerate shell states in the TD. We extend the discussion about identification of superexchange mechanism in the SM\cite{SM}.

The presence of indirect coupling between spins in triangulene spin-1 chains opens a new possibility to control parameters in BLBQ model given by Eq. \ref{eq:BLBQ}. While it might be difficult, or even impossible, to go beyond the Haldane phase on the phase diagram\cite{BLBQphases_Lauchli}, slight variations of $\beta$ might be possible by tuning the energy gap between the inter-triangulene states and the degenerate shell states, which translates into control of the singlet-triplet splitting. The superexchange mechanism discussed in our work for triangulene spin-1 chains should also be responsible for AFM coupling in other triangulene systems, in particular, in recently proposed two-dimensional triangulene crystals \cite{ortiz2022theory,CatarinaTrianguleneCrystals}. Following this path, such newly created AFM spin lattice systems should reveal similarities to antiferromagnetism in transition metal oxides \cite{rao1989transition,Tokura2000TMO,Bhattacharya2014MagneticOxides}, where superexchange mechanism plays a dominant role. Superexchange interaction in anion-bridged dinuclear transition-metal complexes can be described by the Goodenough-Kanamori rules\cite{Goodenoughrules,GOODENOUGH1958287,Kanamori1959SuperExchange}. In analogy to these systems, it might be possible to establish similar rules  for understanding the spin coupling in triangulene spin chains. However, the orbitals originating from the degenerate shell of triangulene are significanlty different than the d-orbitals of the metal ions. Thus, determining if similar rules apply here requires more detailed studies.

\section{acknowledgments}
YS and GB acknowledge support by the Cluster of Excellence “Advanced Imaging of Matter” of the Deutsche Forschungsgemeinschaft (DFG) - EXC 2056-Project 390715994. PP acknowledges support from the Polish National Science Centre based on Decision No. 2021/41/B/ST3/03322. Calculations were partially performed using the Wrocław Center for Networking and Supercomputing WCSS grant No. 317. Y.S. thanks M. Korkusinski, M. Prada for useful discussions.

\bibliographystyle{naturemag}
\bibliography{Fmain}

\end{document}


\graphicspath{}
\preprint{APS/123-QED}

\title{\centering Supplementary materials\\\vspace{0.25em} Superexchange Mechanism in Coupled Triangulenes Forming spin-1 Chains}
\date{12 March, 2024}

\author{Yasser Saleem}
\affiliation{Institut f\"{u}r Physikalische Chemie, Universit\"{a}t Hamburg, Grindelallee 117, D-20146 Hamburg, Germany}

\author{Torben Steenbock}
\affiliation{Institut f\"{u}r Physikalische Chemie, Universit\"{a}t Hamburg, Grindelallee 117, D-20146 Hamburg, Germany}

\author{Emha Riyadhul Jinan Alhadi}
\affiliation{Institute of Physics, Faculty of Physics, Astronomy and Informatics, Nicolaus Copernicus University, Grudziadzka 5, 87-100 Toru\'n, Poland}

\author{Weronika Pasek}
\affiliation{Institute of Physics, Faculty of Physics, Astronomy and Informatics, Nicolaus Copernicus University, Grudziadzka 5, 87-100 Toru\'n, Poland}

\author{Gabriel Bester}
\affiliation{Institut f\"{u}r Physikalische Chemie, Universit\"{a}t Hamburg, Grindelallee 117, D-20146 Hamburg, Germany}

\author{Pawel Potasz}
\affiliation{Institute of Physics, Faculty of Physics, Astronomy and Informatics, Nicolaus Copernicus University, Grudziadzka 5, 87-100 Toru\'n, Poland}

\maketitle


\section{Density functional theory}

All density functional theory (DFT) calculations were carried out with
the {\sc Turbomole 7.5} program package \cite{TURBOMOLE} employing the
B3LYP global hybrid exchange--correlation functional with 20\% of the exact
Hartree-Fock exchange \cite{Dirac1929,Slater1951,Vosko1980,Becke1988,Lee1988,Becke1992}, Ahlrich's triple-zeta split-valence
basis set with polarization functions on all atoms, 
def2-TZVP \cite{Weigend2006}, and the empirical dispersion correction
of Grimme in the third generation \cite{Grimme2010}. Additionally,
we employ the multipole-accelerated resolution of identity approximation 
for Coulomb integrals (MARIJ) \cite{Eichkorn1995,Eichkorn1997,Sierka2003,Weigend2006} to speed-up the calculation of the Coulomb integrals in the self-consistent field (SCF) algorithm.

First, we perform structure optimizations on the spin ground states
approximated by Broken-Symmetry (BS) determinants \cite{Noodleman1981}, 
where the $S=1$ spins of adjacent triangulenes are aligned anti-parallel.
In the structure optimizations convergence criteria of $10^{-7}$~Hartree for 
the energy and $10^{-4}$~Hartree/Bohr for the gradient were employed.
On top of the optimized structures, we perform single-point calculations
for non-spin polarized cations that serve as the reference in the 
full configuration interaction (CI) calculations. The cations are obtained by removing all unpaired
electrons, two for each triangulene subunit, from the systems. In these 
single-point calculations,the same convergence criterion for the energy of $10^{-7}$~Hartree is employed.

\section{Calculation of Coulomb Matrix Elements in the SITE BASIS}
\label{App:CME}
The Coulomb matrix elements $\braket{ij|V|kl}$ are given explicitly as \cite{graphenebook}

\begin{equation}
   \braket{ij|V|kl}= \int\int d\vec{r}_1d\vec{r}_2 \phi^{*}_i(\vec{r}_1) \phi^{*}_j(\vec{r}_2) 
    \frac{2R_y}{\kappa|\vec{r}_1-\vec{r}_2|}\phi_k(\vec{r}_2)   \phi_l(\vec{r}_1) ,
       \label{eq:CME}
    \end{equation}
where, $\vec{r}_1$, $\vec{r}_2$ are the coordinates of electron 1 and electron 2, $\kappa$ is the dielectric constant, and $R_y$ is the Rydberg constant. $\phi_i(\vec{r})$ are $p_z$ Slater orbitals centered on atom $i$, they are given as

\begin{equation}
   \phi_i(\vec{r})=\left(\frac{\xi^5}{32\pi}\right)^\frac{1}{2}ze^{-\frac{\xi}{2}\left|\vec{r}-\vec{r}_i\right|},
       \label{eq:pzorbtial}
    \end{equation}
    where $\vec{r}_i$ is the position of carbon atom $i$, and $\xi = 3.25$ \cite{xipaper}. The lengths are in units of Bohrs. The integrals are solved efficiently in real space using the VEGAS integration alogorithim contained in the GNU scientific library \cite{gsl}. Table~\ref{tbl:coulele} shows all quantum mechanical scattering elements considered and their magnitude for $\kappa=1$. When electrons are far enough away, we take a classical limit of the Coulomb elements, mainly we also take the elements: $\braket{ij|V|ji} = \frac{2R_y}{|\vec{R}_i-\vec{R}_j|}$ for atoms $i$,$j$ beyond NNN. 

\begin{table}[ht]
    \begin{tabular}{c|c}
         Element & Value (eV) \\
         \hline
         $\langle 1, 1 |V| 1,1\rangle$  & $17.311$ \\
         $\langle 1, 2 |V| 2,1\rangle$  & $8.942$ \\
         $\langle 1, 3 |V| 3,1\rangle$  & $5.583$ \\
         $\langle 1, 1 |V| 1,2\rangle$  & $3.027$ \\
         $\langle 1, 2 |V| 3,1\rangle$  & $1.664$ \\
         $\langle 1, 2 |V| 1,2\rangle$  & $0.774$ \\
         $\langle 1, 1 |V| 2,2\rangle$  & $0.774$ \\
         $\langle 2, 2 |V| 1,3\rangle$  & $0.535$ \\
         $\langle 1, 2 |V| 2,3\rangle$  & $0.356$ \\
         $\langle 1, 1 |V| 1,3\rangle$  & $0.306$ \\
         $\langle 1, 1 |V| 1,4\rangle$  & $0.113$ \\
         $\langle 1, 2 |V| 2,4\rangle$  & $0.113$ \\
         $\langle 1, 3 |V| 3,4\rangle$  & $0.122$ \\
    \end{tabular}
    \caption{Table of scattering Coulomb matrix elements of $p_z$ electrons for $\kappa=1$. The numbers 1,2,3, and 4 correspond to atoms where the difference in the numbers corresponds to the neighbor distance i.e. a difference of 1 corresponds to NN, a difference of 2 between numbers corresponds to NNN, and a difference of 3 corresponds to NNNN.}
    \label{tbl:coulele}
\end{table}

\section{Configuration Interaction basis}
The many-body Hilbert space can be divided into smaller subspaces with total spin ${\bf S}$ and azimuthal spin $S^z$. We construct the basis, in the occupation number representation, distributing particles among single-particle states
labeled with spin $S^z$. The total number of possible configurations $N_{\rm cf}$ for $N_{\rm e}$ particles distributed on $N_{\rm st}$ single particle states with a given spin $N_\downarrow$ or $N_\uparrow$, where $N_{\rm e}=N_\downarrow + N_\uparrow$ is determined by a product of binomial coefficients,  $N_{\rm cf}=\binom{N_{\rm st}}{N_{\uparrow}} \cdot \binom{N_{\rm st}}{N_{\uparrow}} $. 
We do not rotate the Hamiltonian matrix to a ${\bf S}$ basis as this is an additional computational cost, and instead determine the ground state from calculations of expectation value of total spin $S$ for each energy eigenstate. 

\section{Triangulene}
We analyze a single triangulene molecule, a triangular graphene quantum dot with zigzag edges shown in Fig.~1(a) of the main text. This quantum dot has broken sublattice symmetry as seen by counting the number of red balls (carbon atoms belonging to sublattice A) and the number of gray balls (carbon atoms belonging to sublattice B). This is also a bipartite lattice, and as such, Lieb's theorem applies\cite{LiebTheorems}. 

\begin{figure}[t]
    \centering
    \includegraphics[width=\linewidth]{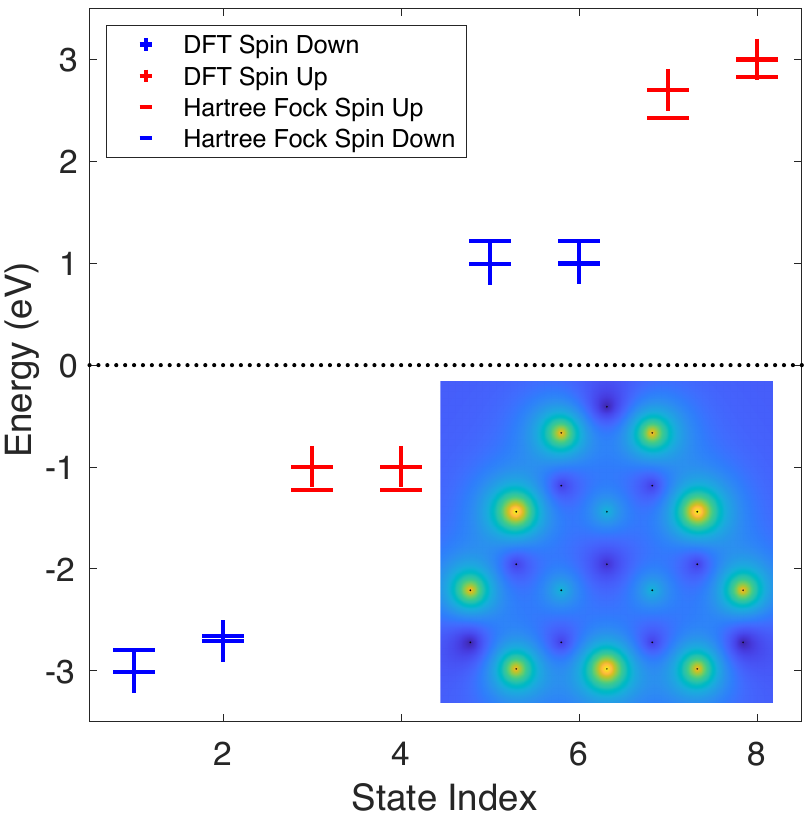}
    \caption{{} Single-particle spectrum of triangulene $N_{\rm Tr}=1$  at half-filling. The horizontal bars are HF results, while the crosses correspond to DFT results. Red color corresponds to spin up levels while blue color corresponds to spin down levels. The inset shows the spin density of the HF ground state.  }
   \label{fig:1triangulenefit}
\end{figure}

We start by performing DFT calculations, and find the ground state to be $S=1$ in agreement with Lieb's theorem and previous experimental and theoretical work \cite{pavlivcek2017synthesis,mishra2019synthesis,ortiz2022theory}. We then perform HF calculations by solving the HF equation. We take $\kappa = 3$ giving the results shown in Fig.~\ref{fig:1triangulenefit}. At the top of the VB there are two degenerate states that are spin up, and a large gap that separates the spin down states. The splitting between the spin up and spin down states arises from a net spin-polarization. These states are localized at the edge, and the spin density of triangulene (shown as an inset in Fig.~\ref{fig:1triangulenefit}) shows the localized spin-1 quasi particle tends to localize at the edge of the triangle.

\section{Comparison between different many-body interacting fermionic models - results for two triangles}
We compare results obtained within different many-body Hamiltonians by restricting Coulomb matrix elements in real space to the dominant ones. Within the Hubbard model, one gets an effective mean-field Hamiltonian given as
\begin{equation}
\begin{split}
    H^H_{MF}&= \sum\limits_{i,l,\sigma}t_{il\sigma} c_{i\sigma}^\dag c_{l\sigma}
   +U\sum\limits_{i,\sigma} 
   \left(\langle \hat{n}_{i,-\sigma}\rangle-\frac{1}{2}\right)\hat{n}_{i\sigma}, 
    \end{split}
     \label{eq:MFHubb} 
    \end{equation}
where the Hubbard parameter $U = \braket{ii|V|ii}$, $\hat{n}_{\rm i}=c^\dag_{i\sigma}c_{i\sigma}$, and  $\langle \hat{n}_{i,-\sigma}\rangle = \rho_{ii,\sigma'}$, are diagonal elements of density matrix with $\sigma'=-\sigma$. For the extended Hubbard model one has
\begin{equation}
\begin{split}
    H^{EH}_{MF}&= \sum\limits_{i,l,\sigma}t_{il\sigma} c_{i\sigma}^\dag c_{l\sigma}
     +U\sum\limits_{i,\sigma}\left(\langle \hat{n}_{i,-\sigma}\rangle-\frac{1}{2}\right)\hat{n}_{i\sigma}\\
   &+\sum\limits_{i\neq j,\sigma,\sigma'}V_{\rm ij}\left(\rho_{jj\sigma'}-\frac{1}{2}\right)\hat{n}_{\rm i\sigma}, 
    \end{split}   
     \label{eq:MFextendHubb} 
    \end{equation}
where $V_{\rm ij} = \braket{ij|V|ji}$.
The Coulomb matrix elements in the basis of mean-field energies $\braket{pq|V|rs}$ are obtained using a basis rotation 
\begin{equation}
    \braket{pq|V|rs}= \sum\limits_{i,j} V_{\rm ij} B^*_{ip}B^*_{jq}B_{jr}B_{is}, 
     \label{eq:CoulElem} 
    \end{equation} 
where $b_{p\sigma} = \sum\limits_{i} B_{ip} c_{i\sigma}$, and $B_{ip}$ are eigenvectors obtained by solving the mean-field Hamiltonian. The form of the many-body Hamiltonian is the same for Hubbard and extended Hubbard, and full interacting models given by Eq. 3 in the main article, but the models differ in the parameters used defined by $\epsilon^{HF}_{p\sigma}$, $\tau_{pq\sigma}$ and Coulomb elements given by Eq. \ref{eq:CoulElem}. A real space spin density distribution for a given many-body state $D$, used when computing the  spin density shown in Fig. 4 in the main article, is calculated using the formula
\begin{equation}
\begin{split}
    \braket{D|\hat{n}_{i\sigma}|D}&= \sum\limits_{p,q} B'^*_{pi}B'_{qi}\braket{D|b^\dag_{p\sigma}b_{q\sigma}|D}\\
    &=\sum\limits_{p,q} B'^*_{pi}B'_{qi}\sum\limits_{c,d}A^*_{c}(D)A_{d}(D) \braket{c|b^\dag_{p\sigma}b_{q\sigma}|d}, 
    \end{split}     
     \label{eq:MBdens} 
    \end{equation} 
where $|D\rangle=\sum\limits_{c}A_{c}|c\rangle$ and $A_{c}$ are expansion coefficients, $|c\rangle = \prod_{p,\sigma} b^\dag_{p\sigma}|0\rangle $ are occupation configuration states and $|0\rangle$ is our vacuum state corresponding to HF state with all valence band states filled (all states below CAS). $B'_{pi}$ are coefficients from inverse transformation, $c_{i\sigma} = \sum\limits_{p} B'_{pi} b_{p\sigma}$, where $B'_{pi}=B^*_{ip}$.

\begin{figure}[t]
    \centering
\includegraphics[width=\linewidth]{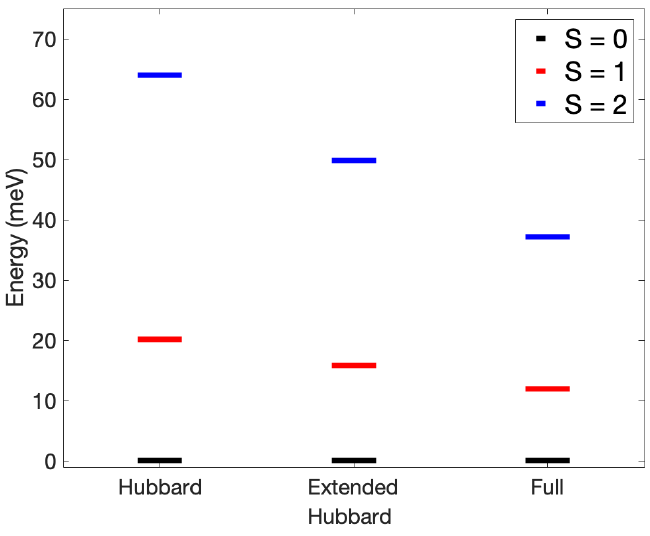}
    \caption{Many-body spectrum obtained for the Hubbard model, extended Hubbard model, and fully interacting model for CAS(6,6) calculations.}
   \label{fig:2trangleMBSpectrumdiffModels}
\end{figure}

Fig.~\ref{fig:2trangleMBSpectrumdiffModels} shows the many-body spectrum for the Hubbard model, extended Hubbard model and the fully interacting model for $N_{\rm Tr}=2$ structure. Calculations in each case were done for CAS(6,6). We find that the Hubbard model tends to overestimate the spin gaps, with the extended Hubbard model capturing most of the quantitative features of the full model. Notice that we are showing only the three lowest energy states, as the excited states are separated from these states by a large gap on the order of hundreds of meV. 

\section{Extended Hubbard model results for $N_{\rm Tr}=4$ triangulenes}
\begin{figure}[t]
    \centering
    \includegraphics[width=\linewidth]{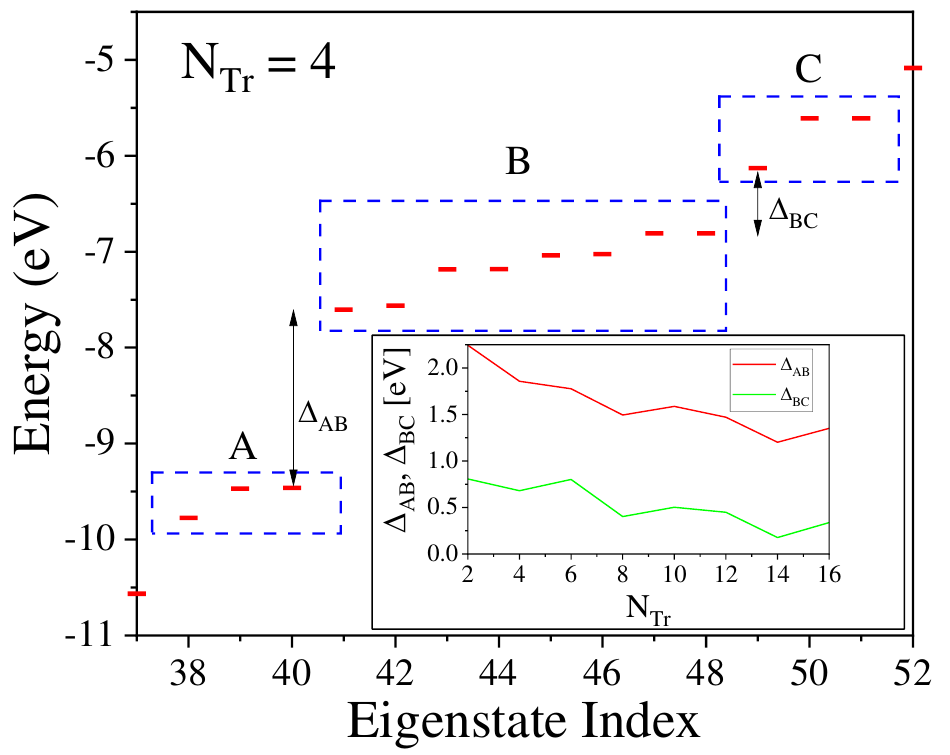}
    \caption{Hartree-Fock spectrum from the extended Hubbard model for $N_{\rm Tr}=4$. Three sets of states, $A$, $B$, and $C$ are indicated, with energy gaps between them, $\Delta_{\rm AB}$ and $\Delta_{\rm BC}$. The inset shows the scaling of the energy gaps with the number of triangulenes $N_{\rm Tr}$ in a chain.}
   \label{fig:extHubbHF}
\end{figure}

Using the extended Hubbard model, we analyze the $N_{\rm Tr}=4$ structure and discuss properties of longer triangulene chains. In Fig. \ref{fig:extHubbHF}, we show the Hartree-Fock energy spectrum indicating the energy gaps between the inter-triangulene states (sectors $B$) and the degenerate shell states (sector $A$ and $C$), $\Delta_{\rm AB}$ and $\Delta_{\rm BC}$. Within this model, the degenerate shell states split into a set of four double degenerate states. There are three inter-triangulene states below and three above the degenerate shell. The energy gap $\Delta_{\rm AB}$ is around twice smaller than $\Delta_{\rm BC}$, and this relation is approximately true also for longer chains, see the inset. These gaps determine the role of excitations and need to be related to scattering Coulomb matrix elements. For $\kappa=3$ used here, we find some of scattering elements between sector $A$ and $B$, and between $B$ and $C$ as large as $0.2$ eV (e.g. $\braket{AA|v|BB}$ and $\braket{AB|v|BC}$). Comparing them to energy gaps $\Delta_{\rm AB}\sim 1.8$ eV, $\Delta_{\rm BC}\sim 0.7$ eV, one can conclude that the states from sector $C$ are as important as the states from sector $A$, see also Section VIII, where we discuss the superexchange mechanism. 

\begin{figure}[t]
    \centering
    \includegraphics[width=\linewidth]{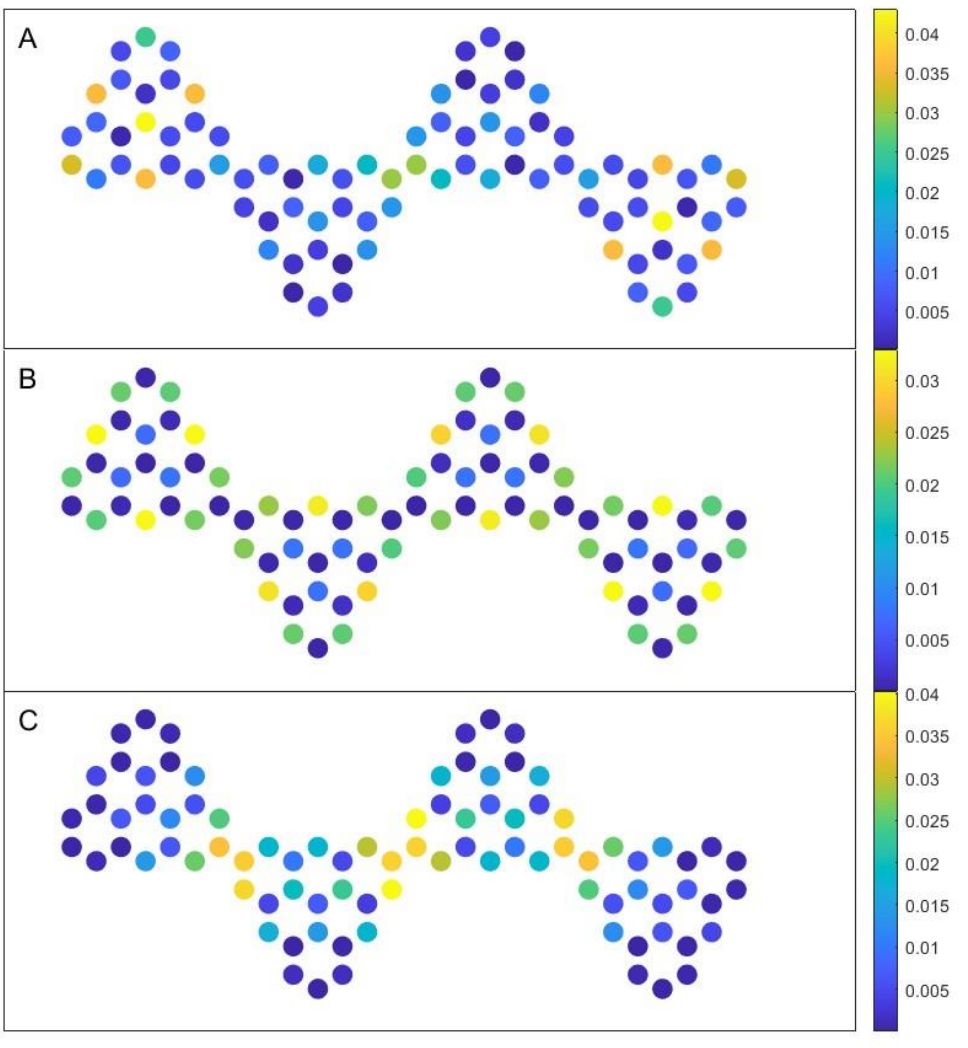}
    \caption{Charge densities of Hartree-Fock spectrum from extended Hubbard model for $N_{\rm Tr}=4$ in three energy sectors indicated in Fig. \ref{fig:extHubbHF}.}
   \label{fig:extHubbHFdens}
\end{figure}

Fig.  \ref{fig:extHubbHFdens} shows charge densities in these three energy sectors; the densities are normalized by the number of states in a given sector. Sector $A$ has localized charge density in the center of the chain, the central inter-triangulene connection and on the first and last triangulenes. On can notice a higher charge density on carbon atoms from the same sublattice as the degenerate shell states, which confirms hybridization with states from sector $B$ (within a tight-binding model, charge density from sector $A$ is similar to that from sector $C$). In sector $B$ charge density is mainly localized on the edges of triangulenes, and only one sublattice. Sector $C$ has charge density on the three connections between the four triangulenes.   

\begin{figure}[t]
    \centering
\includegraphics[width=\linewidth]{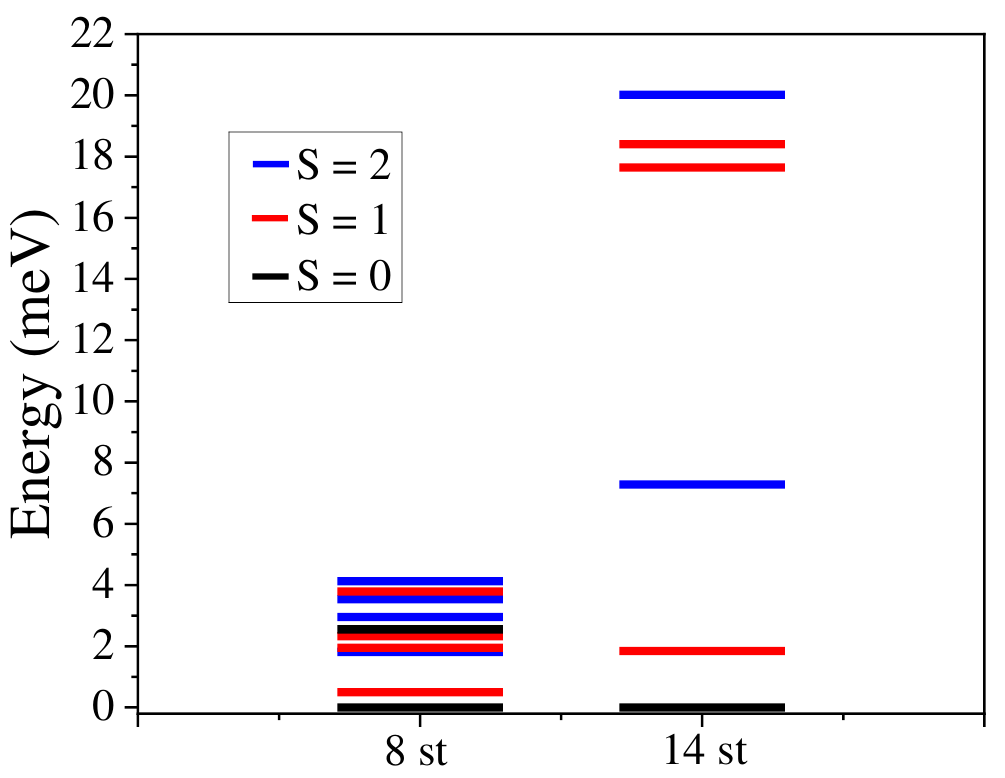}
    \caption{Many-body spectrum from extended Hubbard model for $N_{\rm Tr}=4$ for CAS(8,8) and  CAS(14,14).}
   \label{fig:extHubbMB}
\end{figure}

Fig. \ref{fig:extHubbMB} shows the many-body spectra obtained using the Hamiltonian given by Eq. 3 in the main article with states from only sector $B$ (8 states) and with all three sectors $A$, $B$, and $C$ (14 states). Similar to the $N_{\rm Tr}=2$ case, inclusion of the inter-triangulene states increases the splitting between singlet and triplet, and triplet and quintuplet. One can also see the order of total spin states (we show only the ten lowest energy states for the 8 state calculation) agrees with the order of states in BLBQ model for the lowest five states, compared with Fig. \ref{fig:SpinModels}. The sixth state in the extended Hubbard model calculations with 14 states has $S=2$ (seventh has $S=0$). We attribute these differences between this extended Hubbard model and the BLBQ model to a CAS that is not large enough to converge excited states in this extended Hubbard model calculation.

\section{Spin model results}
\begin{figure}[t]
    \centering
    \includegraphics[width=\linewidth]{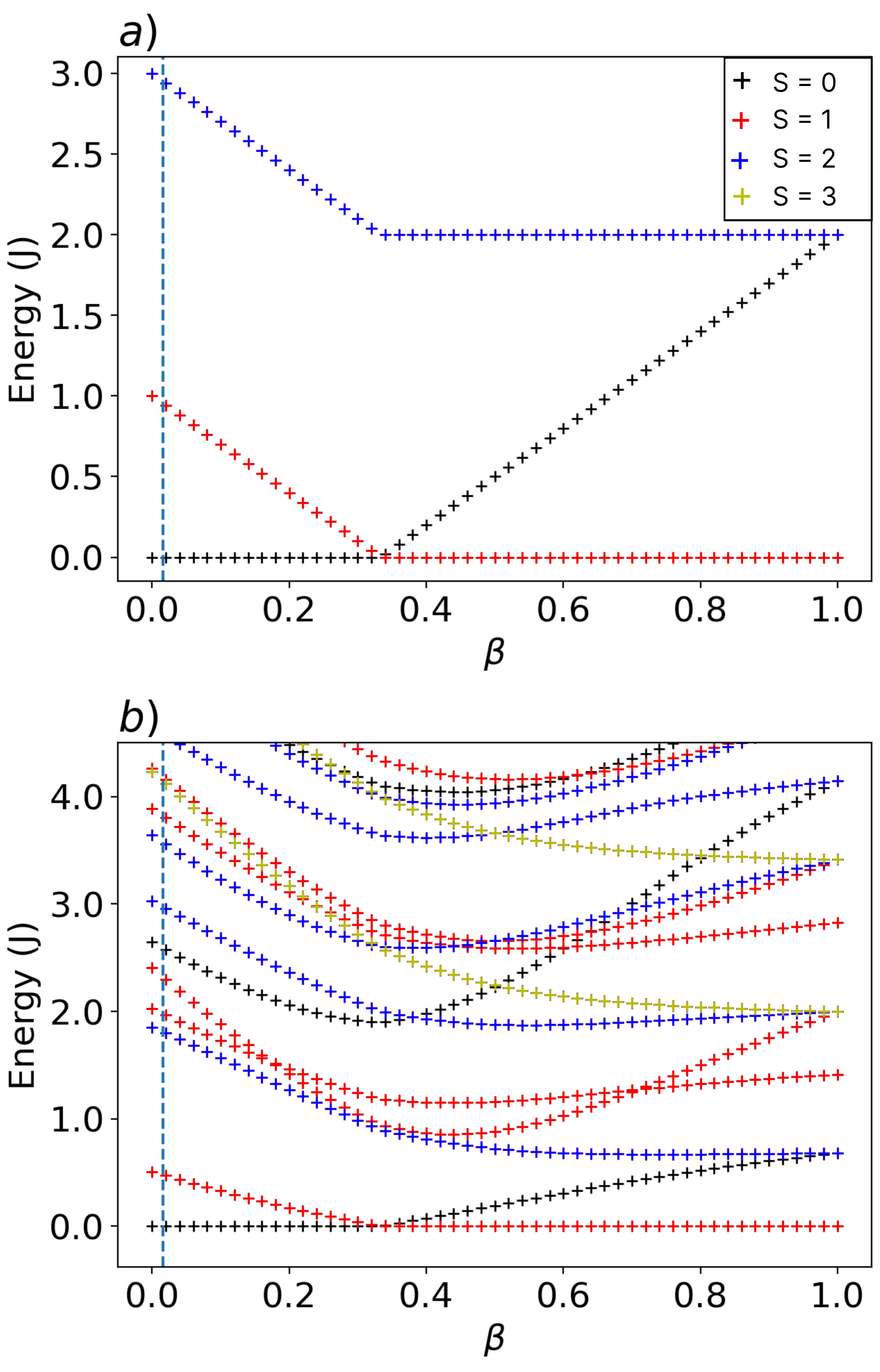}
    \caption{The BLBQ model results as a function of $\beta$ parameter for spin chains with (a) $N=2$  and (b) $N=4$ sites. The blue dashed line indicates the spectra from the Fermionic model in the main text.}
   \label{fig:SpinModels}
\end{figure}
The antiferromagnetic spin-1 Heisenberg model with the added biquadratic operator can effectively describe the chain of triangulane molecules. The effective Hamiltonian is: 
\begin{equation}
    H = J \sum_i [ (\bar{S}_i \cdot \bar{S}_{i+1}) + \beta (\bar{S}_i \cdot \bar{S}_{i+1})^2  ],
\end{equation}
where J is the coupling constant, $\beta$ is the biquadratic term amplitude, and $\bar{S}_i$ is a spin vector of the ith molecule in the chain. 

Using the definition of the ladder operators $\hat{S}_i^\pm = \hat{S}_i^x \pm i\hat{S}_i^y$, one can rewrite the Hamiltonian as:
\begin{equation}
    \begin{split}
        H &= J\sum_i[\frac{1}{2}(\hat{S}_i^+\hat{S}_{i+1}^- + \hat{S}_i^-\hat{S}_{i+1}^+) + \hat{S}_i^z\hat{S}_{i+1}^z \\
        &+\frac{\beta}{4}(\hat{S}_i^+\hat{S}_{i+1}^-\hat{S}_i^+\hat{S}_{i+1}^- + \hat{S}_i^+\hat{S}_{i+1}^-\hat{S}_i^-\hat{S}_{i+1}^+ \\
        &+ \hat{S}_i^-\hat{S}_{i+1}^+ \hat{S}_i^-\hat{S}_{i+1}^+ + \hat{S}_i^-\hat{S}_{i+1}^+ \hat{S}_i^+\hat{S}_{i+1}^- ) \\
        &+ \frac{\beta}{2}(\hat{S}_i^+\hat{S}_{i+1}^-\hat{S}_i^z\hat{S}_{i+1}^z + \hat{S}_i^-\hat{S}_{i+1}^+\hat{S}_i^z\hat{S}_{i+1}^z \\
        &+ \hat{S}_i^z\hat{S}_{i+1}^z \hat{S}_i^+\hat{S}_{i+1}^- + \hat{S}_i^z\hat{S}_{i+1}^z \hat{S}_i^-\hat{S}_{i+1}^+) \\
        &+\beta (\hat{S}_i^z\hat{S}_{i+1}^z)^2 ].
    \end{split}
\end{equation}
The single spin-1 site has three possible $\hat{S}_i^z$  projections: $\ket{\uparrow}=\ket{1}$, $\ket{\downarrow}=\ket{-1}$, and $\ket{0}$. To obtain analytical results, we construct the basis for N=2 chain sites, composed of all combinations of $\hat{S}_i^z$ values. The $\hat{S}^z=0$ subspace is spanned by three basis vectors: $\ket{0, 0}$, $\ket{-1, 1}$, $\ket{1, -1}$. Acting with the Hamiltonian operator on this subspace results in:
\begin{equation}
    \begin{split}
        &\hat{H}\ket{-1, 1} = J(1-\beta)\ket{0, 0} + J\beta\ket{1, -1} + J(2\beta-1)\ket{-1, 1}, \\
        &\hat{H}\ket{0, 0} = 2J\beta\ket{0, 0} + J(1-\beta)\ket{1, -1} + J(1-\beta)\ket{-1, 1}, \\
        &\hat{H}\ket{1, -1} = J(1-\beta)\ket{0, 0} + J(2\beta-1)\ket{1, -1} + J\beta\ket{-1, 1}, \\  
    \end{split}
\end{equation}
which gives us the matrix:
\begin{equation}
    H_{S^z=0}=J\begin{bmatrix}
(2\beta-1) & (1-\beta) & \beta \\
(1-\beta) & 2\beta & (1-\beta) \\
\beta & (1-\beta) & (2\beta-1) 
\end{bmatrix}.
\end{equation}
Similarly, Hamiltonian matrices for other $S^z$ subspaces are:
\begin{equation}
    \begin{split}
        H_{|S^z|=1} &=J\begin{bmatrix}
\beta & 1  \\
1 & \beta 
\end{bmatrix} \\
H_{|S^z|=2} &= J\begin{bmatrix}
    (1+\beta)
\end{bmatrix}.
    \end{split}
\end{equation}
The full Hamiltonian matrix is block diagonal, and its eigenvalues for the singlet, triplet, and quintuplets are  $E(S) = -2J(1-2\beta)$, $E(T)=-J(1-\beta)$, and  $E(Q)=J(1+\beta)$. After shifting eigenvalues by $J(1+\beta)$, we see that only the singlet state depends on the $\beta$:  $E(S) = -3J(1-\beta)$, $E(T)=-2J$, and  $E(Q)=0$. For $\beta < \frac{1}{3}$ the ground state is a singlet and for $\beta > \frac{1}{3}$ it is a triplet state. For $\beta = \frac{1}{3}$, the singlet and triplet states are degenerate, as shown in Fig. \ref{fig:SpinModels}(a). 

Fig. \ref{fig:SpinModels}(b) shows numerical results for a chain of $N=4$ spin-1 sites. The transition between the singlet and triplet ground state again occurs at $\beta=\frac{1}{3}$, but the gap between triplet and quintuplet now depends on $\beta$.



\section{Superexchange mechanism analysis}
\begin{figure}[t]
    \centering
    \includegraphics[width=\linewidth]{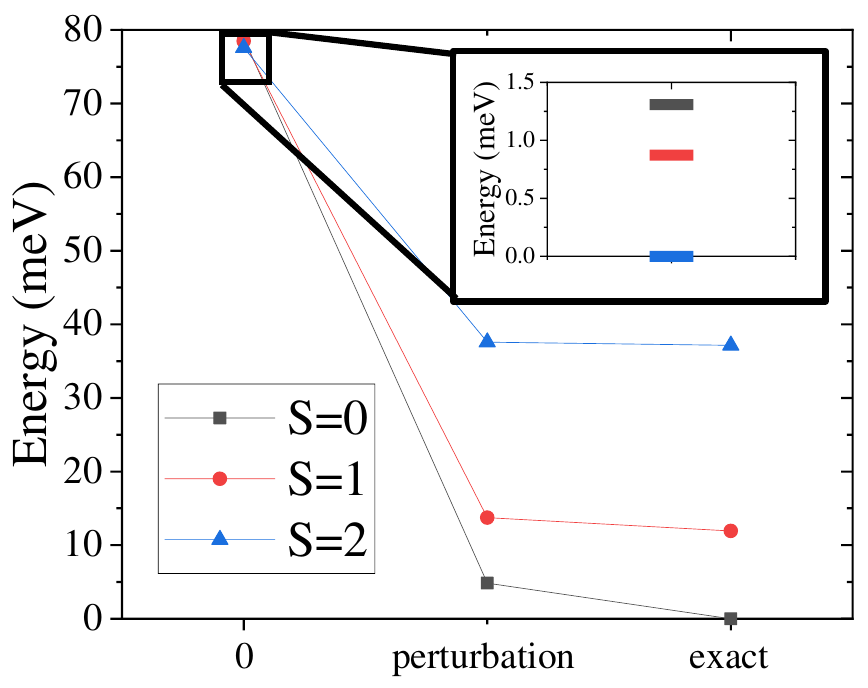}
    \caption{Perturbation analysis of coupling between spins. $0$ indicates the lowest energy total spin states obtained from diagonalization of many-body Hamiltonian matrix within a single occupation subspace of the degenerate shell states. Perturbation indicates energies of these states after energy  corrections included within 2nd order perturbation theory. Exact are energies obtained after diagonalization full many-body Hamiltonian.}
   \label{fig:MBSpectrumdiffModels}
\end{figure}
 Isolated triangulene has a triplet ground state and can be represented by an effective spin-1 site. We analyze the processes responsible for coupling between neighboring spin-1 states on the example of two triangulenes. Our methodology for treating Coulomb interaction relies on a two-step process. First, we include interactions at the Hartree-Fock level for a closed shell system, and next, we populate unoccupied states up to charge neutrality and diagonalize the many-body Hamiltonian within a restricted subspace. For the $N_{\rm Tr}=2$ system, after the HF basis rotation, the degenerate shell states from two triangles form symmetric and antisymmetric linear combinations of states from each triangle, as seen in the wavefunctions of Fig. 3 of the main article. Moreover, we observe strong mixing between the four degenerate shell states and valence and conduction band inter-triangulene states.

This hinders the perturbative analysis. The two isolated triangulenes have four perfectly degenerate edge states, two for each triangulene. The perfect strategy would be to rotate the HF degenerate shell states back to the isolated triangulene basis states. In that case, one could show that coupling between triangulene spin-1 states are through the inter-triangulene states - an indirect superexchange mechanism, one of the main results of this work. However, after HF self-consistent calculations, the degenerate shell states are too strongly hybridized with the inter-triangulene states.  
 
We analyze the superexchange mechanism at the many-body level. We take six HF states to construct our many-body Hilbert space, the highest valence band state, four degenerate shell states and the lowest conduction band state. These states are labelled 5-10 in Fig. 2(c) of the main text, in this analysis, for simplicity we shift the labels to 1-6. The charge neutral system has $N_{\rm el}=6$ electrons. For $S_{\rm z}=0$, one can construct in total, $N_{\rm cf}=400$ configurations. The low-energy subspace, which we call later a single occupation subspace, corresponds to double occupation of the valence band state and single occupation of four degenerate shell states. One can construct six such low-energy configurations. In the occupation representation, these states can be written as 
\begin{equation}
\begin{split}
    |I\rangle &= |1\downarrow, 1\uparrow, 2\downarrow, 3\downarrow, 4\uparrow, 5\uparrow\rangle  \\
    |II\rangle &= |1\downarrow, 1\uparrow, 2\uparrow, 3\uparrow, 4\downarrow, 5\downarrow\rangle  \\
    |III\rangle &= |1\downarrow, 1\uparrow, 2\downarrow, 3\uparrow, 4\downarrow, 5\uparrow\rangle  \\
    |IV\rangle &= |1\downarrow, 1\uparrow, 2\uparrow, 3\downarrow, 4\uparrow, 5\downarrow\rangle  \\
    |V\rangle &= |1\downarrow, 1\uparrow, 2\downarrow, 3\uparrow, 4\uparrow, 5\downarrow\rangle  \\
    |VI\rangle &= |1\downarrow, 1\uparrow, 2\uparrow, 3\downarrow, 4\downarrow, 5\uparrow\rangle, 
    \end{split}
     \label{eq:SingleBasis} 
    \end{equation}
where state $1$ is the valence band state lying just below the four degenerate shell states, states $2-5$ are the four degenerate shell states, and state $6$ is the state just above the four degenerate states (unoccupied here).    
The six configurations defined in Eq. \ref{eq:SingleBasis} can then be rotated into two total spin $S=0$ singlets, three $S=1$ triplets and one $S=2$ quintuplet. 
We obtain the lowest energy state $E_{0}(S)$ within each total spin single occupation subspace by diagonalizing a $2\times 2$ matrix within $S=0$ subspace and a $3\times 3$ matrix within $S=1$ subspace, at the same time appropriately rotating the full Hilbert space within each total spin sector.  This procedure is related to the strong coupling between configurations given by Eq. \ref{eq:SingleBasis}, due to the hybridization between the degenerate shell states and inter-triangulene states.  Next, the obtained lowest energy states within each total spin subspace, are corrected by second order perturbation contributions due to the coupling to the rest of states $E_{i}(S)$ from the many-body Hilbert space (which have been rotated to the total spin subspaces), beyond the single occupation subspace ($S=0$ subspace contains in total $N_{\rm cf}=175$ configurations, $S=1$ contains $N_{\rm cf}=189$ configurations, and $S=2$ contains $N_{\rm cf}=35$ configurations). The perturbative Hamiltonian is written as 
\begin{equation}
E^{(2)}(S)=\sum_{i=1}^{N_{\rm cf}(S)} \frac{|\langle \Psi_0(S)|H|\Psi_i(S)\rangle|^2}{E_{0}(S)-E_{\rm i}(S)},
       \label{eq:perturb}
    \end{equation}    
where $\Psi_0$ and $\Psi_i$ are the many-body wave function corresponding to energy $E_0$ and $E_i$, respectively, and $H$ is the many-body Hamiltonian given by Eq. (3) in the main article. Although this procedure is second order in perturbation theory, if the hybridization between the edge states of the individual triangulenes and the inter-triangulene states were weak, then perturbative treatment of the lowest energy total spin states (corresponding to singlet, triplet and quintuplet) would be appropriate. In that case, fourth order perturbation theory would be required in order to describe coupling of the degenerate states of two triangulenes (and thus the spin-1 quasiparticles) through the inter-triangulene states. 

The energy spectra after diagonalization within the single occupation subspace, and after second order perturbation correction is compared to the exact many-body spectrum in Fig. \ref{fig:MBSpectrumdiffModels}. The order of total spin states after basis rotation within a single occupation subspace is the same as within truncated $N_{\rm st}=4$ states with the $S=2$ quintuplet as the ground state, see Fig. 4(a) in the main article. Coupling of the single occupation subspace to higher energy configurations leads to a change of the order of states that now agrees with predictions within the two spin-1 Heisenberg Hamiltonian, with the singlet as the ground state. Furthermore, results after this coupling, agrees as well as with the experiments~\cite{FaselNaturespinchain}. The energies are close to the exact energies obtained after diagonalization of the full many-body Hamiltonian. We notice that configurations mainly contributing to the perturbation includes all six HF states, thus both the state below and the state above the four degenerate shell states are important, which confirms the existence of an indirect AFM superexchange mechanism.   


\bibliographystyle{naturemag}
\bibliography{SM}